\documentclass[12pt]{iopart}

\usepackage{graphicx}
\usepackage[tight]{subfigure}
\usepackage{iopams}

\begin{document}
\title{Gaussianity revisited: Exploring the Kibble-Zurek mechanism with superconducting rings}

\author{D\ J\ Weir$^1$, R\ Monaco$^2$, V\ P\ Koshelets$^3$, J\ Mygind$^4$ \\ and R\ J\ Rivers$^5$}

\address{$^1$ Helsinki Institute of Physics, Gustaf H\"{a}llstr\"{o}min katu 2a, 00014 Helsinki, Finland}
\address{$^2$ Istituto di Cibernetica del CNR, Comprensorio Olivetti, 80078 Pozzuoli, Italy \\and Facolt$\grave{\rm a}$ di Scienze, Universit$\grave{\rm a}$ di Salerno, 84084 Fisciano, Italy}
\address{$^3$ Kotel'nikov Institute of Radio Engineering and Electronics,
Russian Academy of Science, Mokhovaya 11, Bldg 7, 125009 Moscow, Russia.}
\address{$^4$ Department of Physics, B309, Technical University of
Denmark, DK-2800 Lyngby, Denmark.}
\address{$^5$ Blackett Laboratory, Imperial College London, London
SW7 2AZ, U.K. }

\ead{david.weir@helsinki.fi}

\date{\today }

\begin{abstract}
In this paper we use spontaneous flux production in annular superconductors to shed light on the Kibble-Zurek scenario. In particular, we examine the effects of finite size and external fields, neither of which is directly amenable to the KZ analysis.
Supported by 1D and 3D simulations,
the properties of
a superconducting ring are seen to be well represented by analytic Gaussian approximations which encode the KZ scales indirectly. Experimental results for annuli in the presence of external fields corroborate these
findings.
\end{abstract}


\section{Introduction}

Causality imposes constraints on systems that are strongly out of
equilibrium by restricting the rate of change of correlation lengths
to the relevant causal speed (such as the speed of sound).
The suggestion that causality would constrain correlation lengths was originally made by Kibble in the context of the very early universe~\cite{kibble1976,kibble1980}, and by Zurek for condensed matter systems~\cite{zurek1985,zurek1996}.
Frustration is relieved by the spontaneous creation of topological defects, whose separation reflects the correlation lengths at the time of their appearance; this is known as the Kibble-Zurek (KZ) scenario.

\begin{figure}[tb]
\centering
\includegraphics[width=0.45\textwidth]{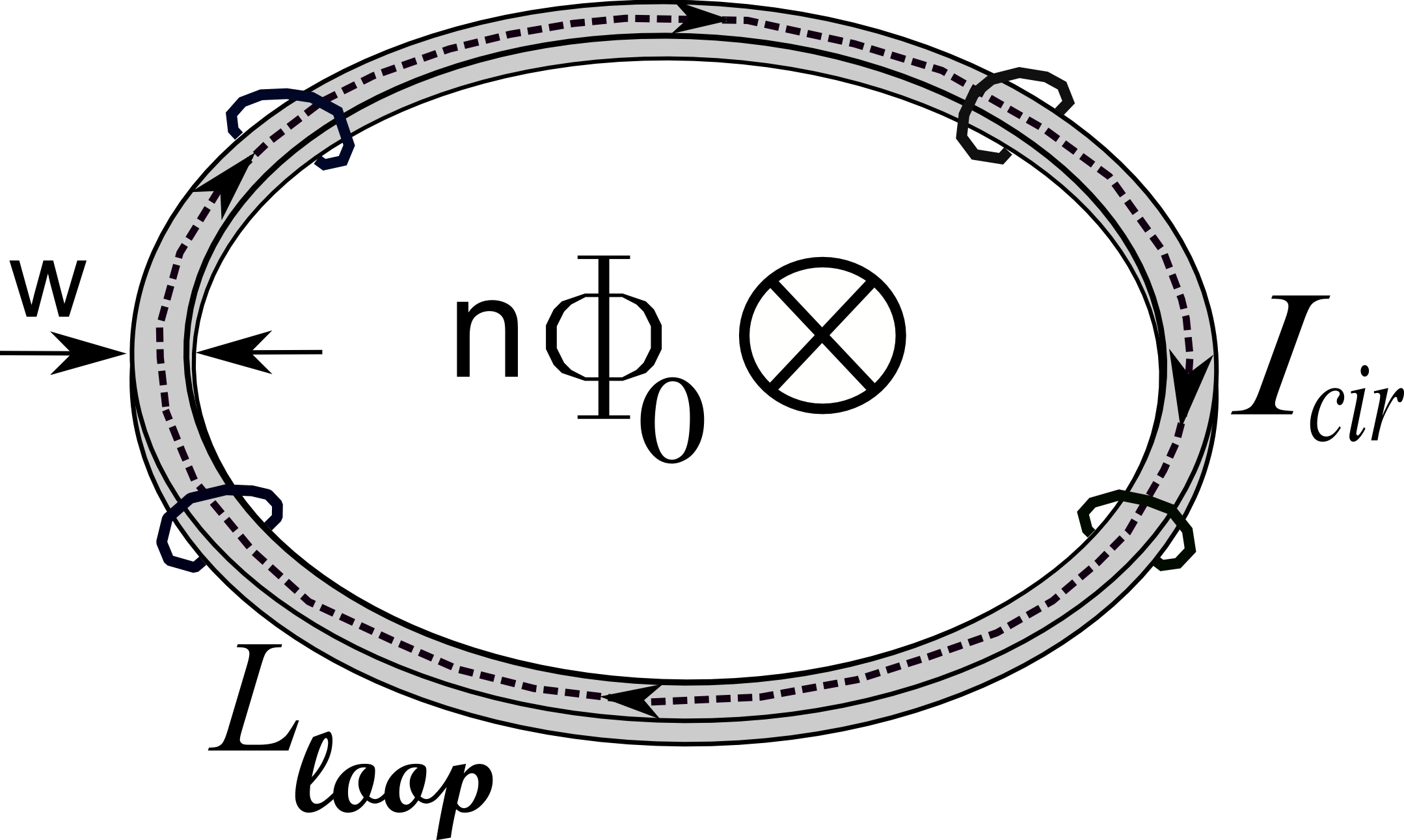}
\caption{3D sketch
of a superconducting
    ring of width $w$ with $n$ trapped magnetic flux quanta, in the
    absence of any external flux.
    A supercurrent $I_{cir}$ circulates around the ring. The magnetic
    field
 lines wrap around the ring section.
\label{ringa}}
\end{figure}

This paper focuses on the behaviour of superconducting annuli.
An idealisation of our
experimental setup -- which we shall discuss later -- is shown in Fig.~\ref{ringa}.
To model the system we assume a simple complex scalar field $\phi(x)$ that can act as a proxy for a Cooper pair, decomposed as
\begin{equation}
\phi(x) = |\phi(x)| e^{i\theta(x)}.
\end{equation}
Initially we restrict our analysis to the inner radius of the annulus, where $x$ measures the distance along the ring, circumference $C$.
On quenching from the normal to the superconducting phase, we can define a winding number density $n(x) = \partial_x\theta(x)/2\pi$ along this circumference.
In the absence of any external fields the total winding number $n$ around the loop
is zero on average, but will have non-zero variance
\begin{equation}
\langle n^2\rangle = \int_0^C dx\int_0^C dy\;\left< n(x)\, n(y)\right>,
\label{Dn2}
\end{equation}
 and it is this that we measure, indirectly, through the matching flux generated in the interior of the annulus, which is directly observable. This flux is quantised (as fluxoids) in units of $\Phi_0 = hc/2e$.

Suppose the correlator is described through the single length scale ${\bar\xi}$. For large annuli, for which $C\gg{\bar\xi}$, we can replace Eq.~(\ref{Dn2}) by
\begin{equation}
\langle n^2\rangle = \frac{C}{2}\int_0^C dx\;\left< n(x)\, n(0) \right>\approx \frac{C}{2}\int_0^{\infty} dx\;\left<n(x)\,n(0)\right>,
\label{dn2}
\end{equation}
if we assume that the two-point correlation function
$\left< n(x)\, n(y)\right>$ is of short range compared to $C$.
 Simple dimensional analysis then gives
\begin{equation}
\langle n^2\rangle = a C/\bar\xi,
\label{per0}
\end{equation}
the perimeter law assumed in the Kibble-Zurek picture, interpreted as a random walk in phase along the circumference in steps of length $\bar\xi$.

To estimate the length scale ${\bar\xi}$
we first repeat the Kibble-Zurek argument of Refs.~\cite{zurek1985,zurek1996}, assuming rapid cooling through a continuous transition with transition temperature $T_c$. Suppose $\xi (t)$ is the {\it adiabatic} correlation length for $n(x)$ at time $t$, diverging at the transition at $t = 0$ as the temperature $T(t)$ changes $(T(0) = T_c$).
 If the system is initially homogeneous and isotropic, the KZ scenario proposes that on approaching the transition the increasing correlation length, originally changing adiabatically, will freeze (the `impulse' regime) at time $t_< < 0$, when $\dot\xi (t_<)\approx c(t_<)$, where $c(t)$ is the relevant causal speed at time $t$.  Equivalently, this can be rephrased in terms of critical slowing down~\cite{zurek1985,zurek1996}.

An alternative approach suggests that, {\it after} passing through the transition, the system will unfreeze into a new adiabatic regime at a time $t_> > 0$ when $|\dot\xi (t_>)|\approx c(t_>)$, and that it is the distance scales here that set the domain size. This gives us the same scaling behaviour and, indeed, it was proposed in the early literature that $t_<$ is as good a time as $t_>$ for estimating domain size~\cite{kibble1980,zurek1985}. This shift of viewpoint -- that the relevant dynamics for quantitative behaviour occur after the transition rather than before -- is supported by numerical simulations~\cite{antunes2006}.

The correlation length $\bar{\xi}$ that sets the scale for phase change along the ring in Eq.~(\ref{per0}) is estimated to be
\begin{equation}
\bar{\xi}\approx \xi(t_>) = \xi_0 \left(\frac{\tau_\mathrm{Q}}{\tau_0}\right)^\sigma,
\label{xibar}
\end{equation}
where $\xi_0$ and $\tau_0$ are system-dependent and $\tau_Q$ is the quench time (the inverse quench rate through the transition).
For Type-II superconductors $t_> = \sqrt{\tau_0\tau_Q}$ and $\sigma = 1/4$ in the mean-field approximation.

Such scaling behaviour 
can have a different origin than simple causal unfreezing. Defect formation may be thought of as arising from the growth of unstable long wavelength modes -- as the field rolls off the potential `hill' --
rather than through direct causal bounds. This is supported indirectly by the many simulations (e.g.~\cite{zurek2,zurek3,bettencourt}) demonstrating the validity of
Eq.~(\ref{xibar}) based on {\it time-dependent} Ginzburg-Landau (TDGL) theory and we shall give an explicit demonstration later.

This shift of view is crucial when considering laboratory systems. Whereas the early universe can be taken as homogeneous to a very high degree, the same cannot be said of experiments.
One step to limit inhomogeneity is to make the system small. However, this
incurs finite size effects.
Specifically, the scaling behaviour of Eq.~(\ref{xibar})  assumes that $C\gg \bar\xi$. However, if we think in terms of growth of mode amplitudes {\it after} the quench there are always modes comparable in wavelength to the system size; their growth can still contribute to defect formation.

The KZ length $\bar\xi$ can of course be recovered from a picture of expanding mode amplitudes for {\it large} systems. If $t_>$ is a guide for the time at which defects form, then it is much smaller than the time taken for the order parameter field(s) to experience the degenerate vacua. If they remain so close to the unstable `vacuum' that the non-linearities of the backreaction can be approximately ignored, the effectively linearised field equations give rise to Gaussian field correlation functions. These linearised equations generally lead to KZ scaling behaviour for fast quenches in large systems,  both non-relativistic and relativistic \cite{karra1997,karra1998,moro}.
More recently, preliminary simulations which further support underlying Gaussian behaviour explicitly for superconductors were performed by us in Refs.~\cite{weir2011,weir2012}, and this paper builds on this analysis.

For {\it homogeneous} small systems with $C < \bar\xi$,  which are inaccessible to the KZ picture, the Gaussian approximation permits an analytic replacement of Eqs.~(\ref{per0}) and (\ref{xibar}) which encodes the KZ scale $\bar\xi$ despite the smallness of the system.
 Furthermore, the same Gaussian approximation is applicable to explicit symmetry breaking driven by the application of external fields, for which the KZ scenario again offers little help. This is an interesting problem in its own right and is a serious problem because of the presence of stray fields in the laboratory. Scaling behaviour of the form of Eq.~(\ref{xibar}) can only be demonstrated experimentally once external fields have been countered.

This paper is organised as follows. For narrow rings we might hope that the system behaves one-dimensionally; in the next section we reexamine the Gaussian approximation (or rather, two
related Gaussian approximations) both for large and small idealised 1D systems.
 We then show that numerical simulations provide strong support for Gaussian behaviour, both in the power spectrum of field ordering and in the damping of fluxoid generation in small systems.
In particular, we explore the way that the power of field fluctuations is driven into long wavelengths through the growth of unstable modes, in good accord with our Gaussian expectations.
We conclude with a discussion of new experimental results for fluxoid production in an external field and show that it too, is explicable in terms of Gaussian fluctuations. Good agreement with 3D simulations is observed.

\section{Gaussian behaviour for 1D systems}

Unsurprisingly, 1D systems are the easiest for which we can make analytic predictions and implement numerical simulations.
There are two strongly related variants of Gaussian approximation that we shall consider; each has its strengths.

 \subsection{Long superconductors: Gaussian probabilities}

 The first Gaussian approximation, proposed in Refs.~\cite{monaco2006b,monaco2008r}, is predicated on the perimeter rule, Eq.~(\ref{per0}).
We begin by considering a large loop, divided up into $N$ domains, in each of which $\theta$ is a constant. We assume that there is no correlation between the values of $\theta$ in adjacent domains but that the shortest path in phase (the geodesic rule) will be taken when jumping from one domain to the other. Let $G_N(\Delta\theta)$ be the probability that the change in phase $\theta$ is $\Delta\theta$ after the $N$ links that make the loop. The assumed lack of correlation means that the probability of ending with a  phase shift of $2\pi m$ (net fluxoid number $m$) is:

\begin{equation}
p_m(N) = \int_{-\pi + 2m\pi}^{\pi + 2m\pi}d\Theta\,G_N(\Theta).
\label{fmN}
\end{equation}
To bring this into correspondence with the KZ scenario, we should identify the domain size as comparable to $C/{\bar\xi}$, i.e., $C = aN{\bar\xi}$, where $a = O(1)$.
With this in mind,  we assume that the total phase change $\Delta\theta$ around the loop can be expressed as the sum of a random term $\Theta$ and a geodesic-rule correction $\delta\Theta$, necessary to obtain an integer winding number~\cite{monaco2008}. It is convenient to relax $N$ to be a {\it continuous} variable. We further assume that $\Theta$ has a normal distribution with average $\bar \Theta = 0$ and variance $\sigma^2$
proportional to $N$ i.e.,

\begin{equation}
G_N(\Theta) = \frac{1}{\sqrt{2 \pi \sigma^2(N)} }
\exp- {\frac{\Theta^2}{2 \sigma^2(N)}}.
\label{Ggauss}
\end{equation}
The trapping probabilities $p_m = p_{-m}$ are easily found:
In particular, since in our experiments we rarely see more than one fluxoid, we are primarily interested in
\begin{equation}
 p_0(N) =
\textrm{erf}\bigg[ \frac{\pi}{\sqrt{2 \sigma^2(N)}}\bigg]
\label{fp10}
\end{equation}
and
\begin{equation}
{\bar p}_1 = p_1(N)+ p_{-1}(N) =
\bigg[\textrm{erf} \frac{3\pi}{\sqrt{2 \sigma^2(N)}} - \textrm{erf} \frac{\pi}{\sqrt{2 \sigma^2(N)}}\bigg].
\label{fp11}
\end{equation}

From the large-N behaviour of the $p_m$ it follows that, for large rings,
$\sigma^2(N) = 4\pi^2 \langle n^2\rangle$. Also, we can identify Eq.~(\ref{Ggauss}) from the central limit distribution by taking $\sigma^2 (N) = \frac{\pi^2}{3} N$ for large $N$. Further details may be found in Ref.~\cite{monaco2008}.

\subsection{A Gaussian variant: Gaussian distributions}

\begin{figure}[tb]
\centering
\includegraphics[width=0.75\textwidth]{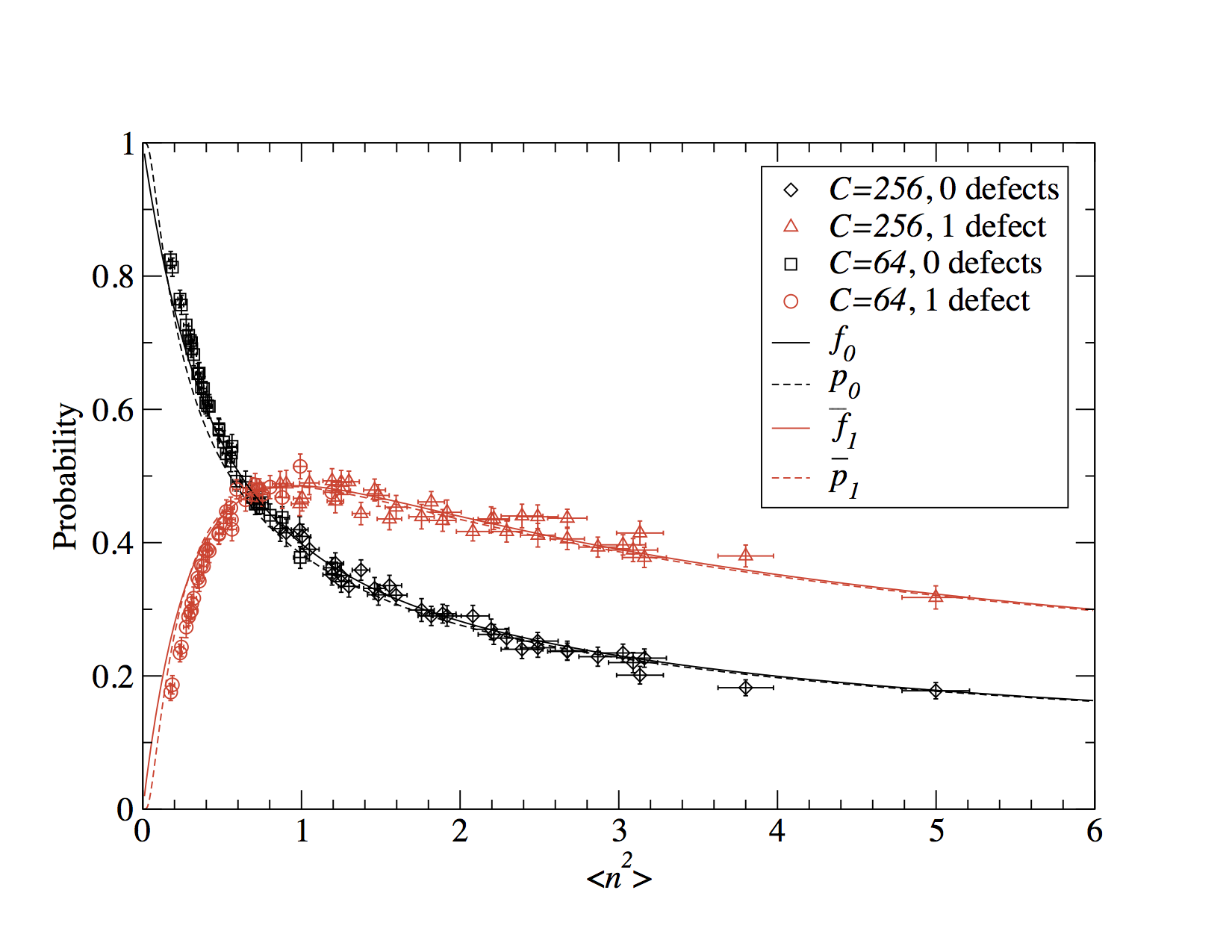}
\caption{We plot ($p_0$, $f_0$) and then ($\bar p_1$, $\bar f_1$) against $\langle n^2\rangle$ for a  1D superconducting ring to show the similarity of the Gaussian approximations. In particular, both $\bar p_1$ and $\bar f_1$ have maximum values of approximately $0.49$. Both are contrasted with the results of the 1D simulation.
Because neither approximation takes periodicity into account the likelihood of seeing flux at low levels is overestimated; see Fig.~\ref{fig:volexpsup}. Otherwise agreement is good within errors.\label{pfs}}
\end{figure}

For our second approximation, rather than assume the Gaussian distribution for $\Theta$ of Eq.~(\ref{Ggauss}), we assume that the winding number density field $n(x)$ is a Gaussian field~\cite{monaco2008}. It then follows that the all-important probabilities of finding no fluxoids or one fluxoid are
 \begin{eqnarray}
f_0 &=& \frac{1}{2\pi}\int^{\pi}_{-\pi}dz\,\exp
(-z^2\langle n^2\rangle /2)\label{fs0}
 \end{eqnarray}
 and
  \begin{eqnarray}
{\bar f}_1 = f_1 + f_{-1} &=& \frac{1}{\pi}\int^{\pi}_{-\pi}dz\,\exp
(-z^2\langle n^2\rangle /2)\cos z\label{fs1}
 \end{eqnarray}
 respectively.
 We use the notation $f_m$ for the probability of winding number $m$ in this approximation to distinguish it from the $p_m$ of the previous section for Gaussian winding number densities.
None of equations (\ref{fp10}) to (\ref{fs1}) take periodicity into account.
In Fig.~\ref{pfs} we show how the predictions for $p_n$ and $f_n$ are
almost indistinguishable, only differing slightly at very small
probabilities where periodicity is important.

The Gaussian approximations so far do not give the KZ scaling behaviour of Eq.~(\ref{xibar}). To see this we make the further Gaussian approximation that, on decomposing the complex order parameter field as $\phi = (\phi_1 + i\phi_2)/\sqrt{2}$, $\phi_1$ and $\phi_2$ are independent Gaussian fields.
The correlation function for the winding number density is now determined by the correlation function $G(x)$ for the field components, defined by
\begin{eqnarray}
\left<\phi_{a}(x)\,\phi_{b}(y)\right>
=\delta_{ab}G(|x-y|).
\label{eq:twentyfour}
\end{eqnarray}

We now use the fact that, for damped systems, the Gaussian approximation corresponds to a linearisation of the TDGL equations. Periodicity is unimportant for large rings and we find (see Ref.~\cite{rivers2001}) that the field correlation function $G(r)$ at the defect formation time for large systems takes the form
\begin{equation}
G(r,t) = \int \frac{dk}{2\pi} \, e^{i k x } P(k, t) \label{diag}
\end{equation}
in which the power spectrum $P(k,t)$ has a representation in terms
of the Schwinger proper-time $\tau$ (in the dimensionless units of $\tau_0 = \xi_0 = 1$) as (approximately)~\cite{karra1997,karra1998, rivers2001}
\begin{equation}
P(k, t) \propto \int_{0}^{\infty} d\tau \,e^{-\tau
k^{2}}\,e^{-\int_{0}^{\tau} ds\,\,\epsilon (t- s/2)}
\label{lgpower}
\end{equation}
where  $\epsilon (t)\equiv T(t)/T_c$ is the reduced temperature. We assume the dependence $\epsilon (t) = -t/\tau_Q$ for a quench linear in time in the vicinity of the phase transition where fluxoid formation takes place. Adopting this $\epsilon(t)$ gives
\begin{equation}
P(k, t) \propto \int_{0}^{\infty} d\tau \,e^{-\tau
k^{2}}\,e^{t\tau/\tau_Q}\,e^{-\tau^2/4\tau_Q}
\label{lgpower2}
\end{equation}
$P(k) \propto k^{-2}$ for large $k^2$.
In the dimensionless units above, the time for the formation of fluxoids is $O(t_>) = O( \sqrt{\tau_Q})$.
When $D\geq 2$ the integral is dominated by the ultraviolet for slow quenches, which needs to be cut off at distance scale $\xi_0$.

It remains to convert the Gaussian behaviour of the fields into the Gaussian behaviour of the winding number.
We observe that
\begin{eqnarray}
n(x)\equiv\frac{1}{\left|\phi(x)^{2}\right|}\overline{n}(x)
=\frac{1}{\left|\phi(x)^{2}\right|}\left(\phi_{1}\partial_{x}\phi_{2}-\phi_{2}\partial_{x}\phi_{1}\right)
\label{eq:six}
\end{eqnarray}
A simple approximation, sufficient for our purposes, is to take~\cite{swarup}
\begin{eqnarray}
\left\langle n\left(x\right) n\left(0\right)\right\rangle
&\approx& \frac{\left< {\bar n}(x) \, {\bar n}(0)\right>}{\left< |\phi^2|\right>^2}
=\frac{2}{(2\pi)^2}\left[f'\left(x\right)^{2}-f''\left(x\right)f\left(x\right)\right]
\label{napprox}
\end{eqnarray}
where
$f\left(x\right) = G(x)/G(0)$. Provided quenches are fast enough that we can ignore the ultraviolet effects,
it follows from Eq.~(\ref{lgpower2}) that $f$
takes the Gaussian form
\begin{equation}
 f(r)\approx e^{-r^2/2\xi_r^2}
 \label{fshort}
\end{equation}
for {\it small} $r=|x|$.
Note that this is evaluated at the KZ time $t_> \approx \tau_Q^{1/2}$, hence we have $\xi_r\approx {\bar\xi}$, the KZ separation length~\cite{karra1997,karra1998}. This is appropriate for a damped system at {\it short} distance.

This creation of winding number can be thought of as due to the unstable long wavelength modes of the field growing as fast as possible, with no backreaction in the short time necessary. It follows that, on restoring dimensional units,
\begin{equation}
\langle n^2\rangle \approx a\frac{C}{\xi_0}\bigg(\frac{\tau_Q}{\tau_0}\bigg)^{-\sigma}.
\label{bign2}
\end{equation}
All the above continues to ignore periodicity. The advantage of choosing Gaussian \textsl{fields} is that they permit the imposition of periodic boundary conditions in a way that the Gaussian distributions described above do not.

\subsection{Small annuli: Gaussian approximation}

We would expect that small systems, whether superconductors or not, would show proportionately less defect production per unit length than larger systems because of end effects or periodicity. Let us consider small annuli for which $C/{\bar\xi}\ll 1$.

The periodicity of $f(x)$ in $x$ (mod $C$) is now important.  The effect of periodicity is to discretise $k$ in Eq.~(\ref{lgpower2}). To implement this in the approximation of Eq.~(\ref{fshort}), we need to replace $f(x)$  by its periodic generalisation, the Jacobi $\vartheta$ function~\cite{weir2011}
\begin{eqnarray}
 f(x)_{per} &=& \frac{\vartheta_3(\pi x/C|2\pi i\xi_r^2/C^2)}{\vartheta_3(0|2\pi i\xi_r^2/C^2)}
 \\
& \approx& 1- 4\sin^2(\pi x/C)~e^{-2\pi^2\xi_r^2/C^2},
\label{JTF}
\end{eqnarray}
whence, from Eq.~(\ref{Dn2}),
\begin{equation}
\langle n^2 \rangle = {\cal O}(e^{-4\pi^2\xi_r^2/C^2}).
\end{equation}
Thus, rather than the power falloff for large loops we have exponential damping
\begin{eqnarray}
\ln \langle n^2 \rangle &\approx& -4\pi^2\xi_r^2/C^2 + \mathrm{ const.}
\label{smallandslow}
\end{eqnarray}
Since it is small, $\langle n^2 \rangle\approx f_1$, the probability of finding single winding number along the annulus.
In principle $\xi_r\propto \tau_Q^{\sigma}$, but some caution is necessary in that the approximation of Eq.~(\ref{napprox}) is too simple,  which makes the single term in Eq.~(\ref{smallandslow}) only approximate, although it preserves the correct features~\cite{swarup}. In Ref.~\cite{weir2011} we showed that exponential damping does take place. However, we can go further in that, if we combine Eq.~(\ref{bign2}) for $\sigma = 1/4$ with Eq.~(\ref{smallandslow}), then in units $\xi_0 = \tau_0 =1$
\begin{equation}
\langle n^2 \rangle = F(\tau_Q/C^4)
\label{uni}
\end{equation}
for some function $F$ interpolating between power behaviour and exponential damping that covers all sized annuli and quench rates.

\subsection{Numerical simulations}
\label{sec:numerical}

We simulate our 1D superconductor with the $\mathrm{U}(1)$ scalar field theory of the complex order parameter $\phi$, given previously, on a ring of circumference $C$ with periodic boundary conditions.
To mimic a temperature quench through its critical point at time $t=0$ we take an explicitly time-dependent potential
\begin{equation}
\label{eq:pot}
V(|\phi|^2) = \epsilon(t) |\phi|^2 + \frac{1}{2} b |\phi|^4.
\end{equation}
Rather than just take $\epsilon$ linear in $t$, as we did in Eq.~(\ref{lgpower2}), we now adopt the more realistic behaviour
\begin{equation}
\label{eq:quench}
\epsilon(t) =\left\{\begin{array}{cl}
1, & \qquad t < - \tau_\mathrm{Q} \\
 - t/\tau_\mathrm{Q}, & \qquad  -\tau_\mathrm{Q} < t< \tau_\mathrm{Q} \\
- 1, & \qquad t > \tau_\mathrm{Q}
   \end{array}\right. ,
   \label{epsilon}
\end{equation}
to model a slow quench.
We start at $t=-2\tau_Q$ and continue until
$t=4\tau_Q$, by which time the defects have frozen out. This system is
modelled by a damped second-order Langevin equation
with zero mean Gaussian noise.
Our linearised Gaussian approximations are independent of $b$ in the first instance; we have performed our simulations for varying $b$ without any noticeably different results.
In evolving the equations a stochastic leapfrog method is
used~\cite{Borrill:1996uq}. Further details may be found in
Ref.~\cite{weir2011}.

Our first application of this simple simulation
is to test the validity of our Gaussian approximations discussed earlier. In Fig.~\ref{pfs} we show the observed frequencies for trapping a given number of fluxoids as a function of $\langle n^2\rangle$. This is compared with the Gaussian predictions $(p_0, f_0)$ and $({\bar p}_1, {\bar f}_1)$. Agreement is surprisingly good.  In particular, the Gaussian upper bound on finding a single fluxoid at about $0.49$ that follows from Eqs.~(\ref{fs1}) and (\ref{fp11}) is saturated in the simulation (within errors). This supports the Gaussian approximation in its simplest guise.

\begin{figure}[tb]
\centering
\subfigure[ ]{\includegraphics[width=0.49\textwidth]{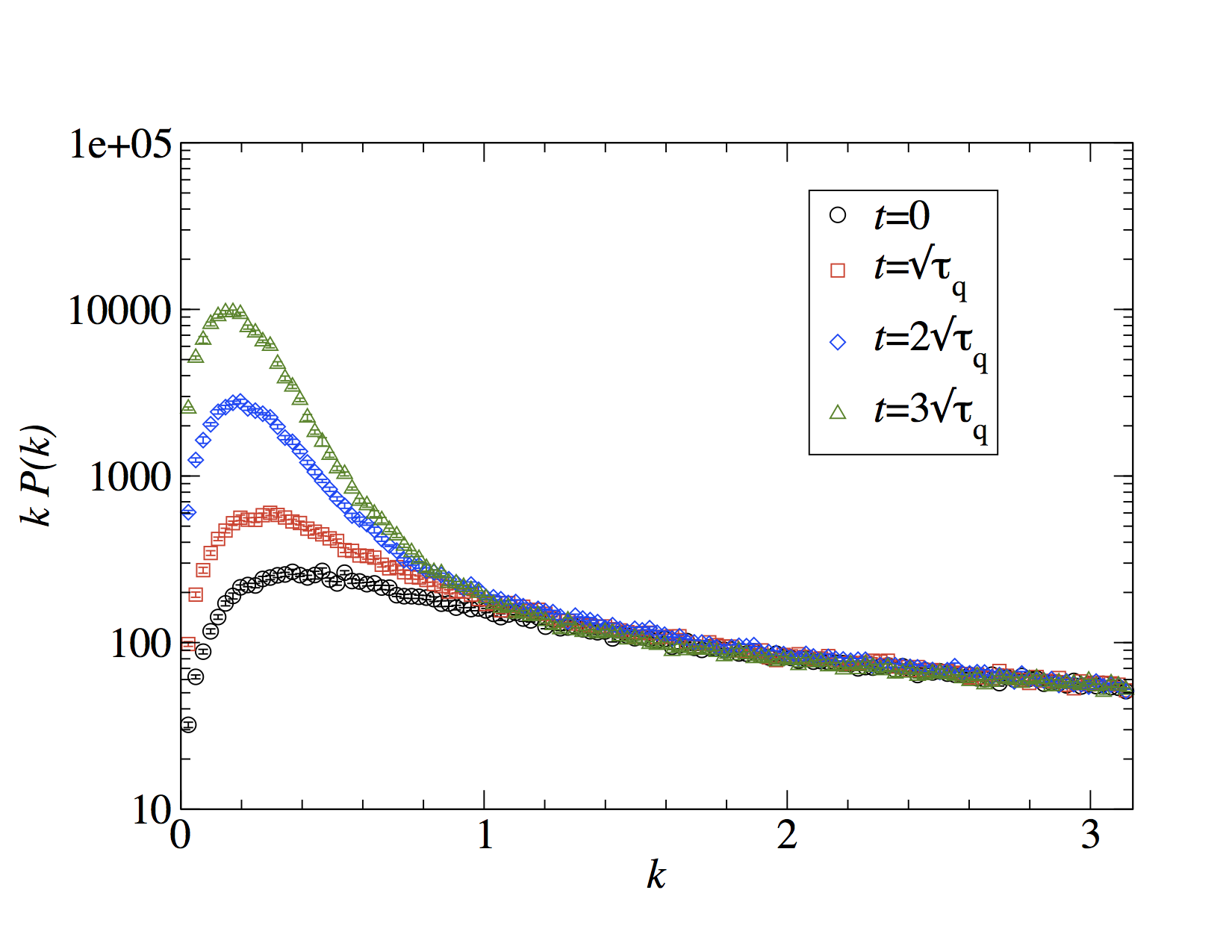}}
\subfigure[ ]{\includegraphics[width=0.49\textwidth]{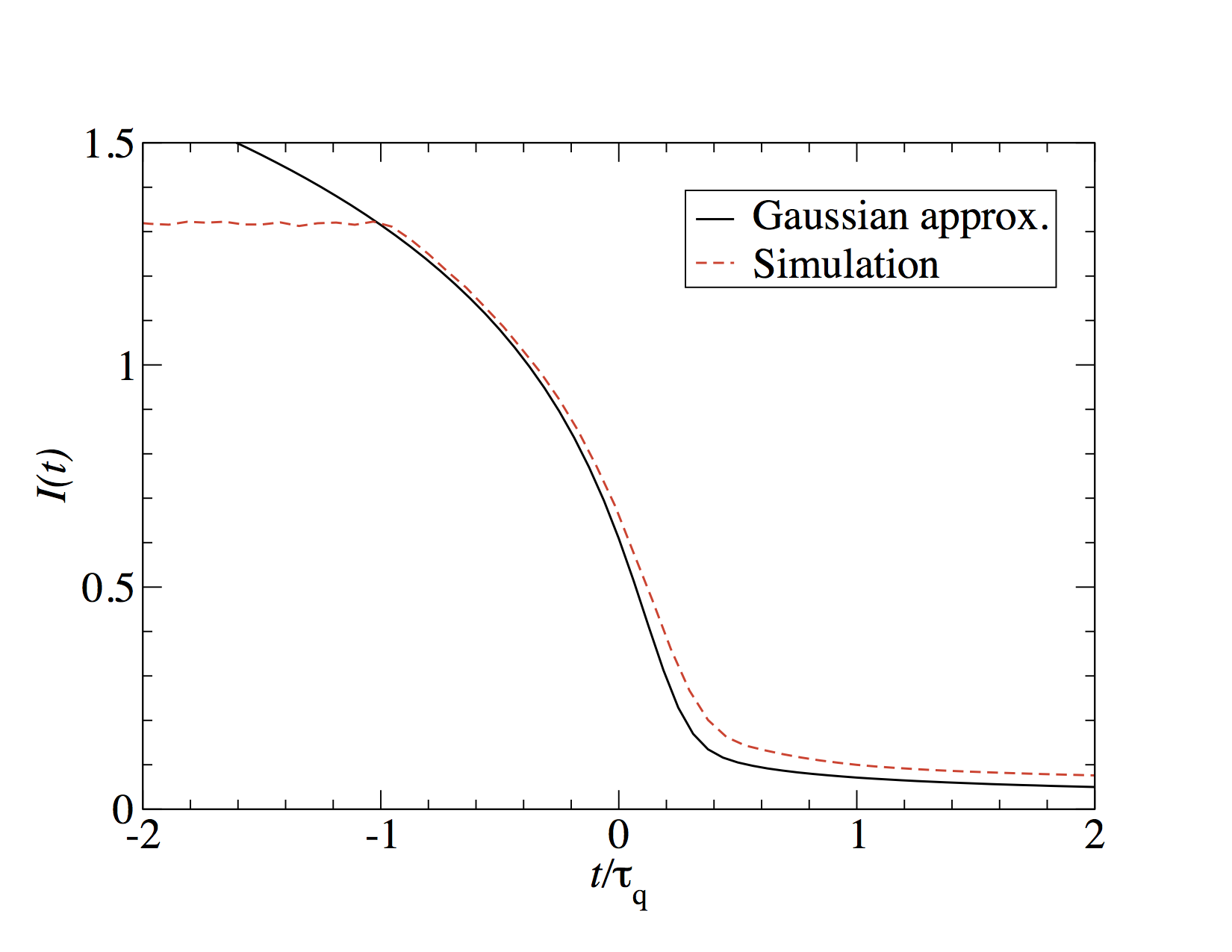}}
\caption{(a) We plot $kP(k)$ against $k$ for a variety of $t > 0$, in multiples of the KZ time $t_> = \sqrt{\tau_Q}$ for $\tau_Q = 32$ on a lattice of size $C = 256$. The characteristic wavelength increases with increasing time as the field orders itself.
(b) We plot the first moment of $P(k)$ from the simulation and compare it to its value from (\ref{lgpower2}).
An identifiable scaling length is only established for $t \gtrsim 3 t_>$. This is reasonable from the KZ viewpoint since $t_>$ is the {\it minimum} time at which the system could unfreeze.\label{kP}}
\end{figure}

 We can say more. The analysis that led to Eqs.~(\ref{lgpower}) and (\ref{lgpower2}) for the power spectrum of the fluctuations is dimension independent and only relies on the Gaussian nature of the fields.  As the fields become ordered the power is driven into longer wavelengths. This can be seen most simply by looking at moments of $P(k)$, for which we expect a peak at ${\bar k}\approx {\bar\xi}^{-1}$.
As an example, in Fig.~\ref{kP} we take the first moment
\begin{equation}
I(t) = \frac{\int_0^{2\pi}dk~kP(k)}{\int_0^{2\pi}dk~P(k)}
\end{equation}
for a given $\tau_Q$, since it is less susceptible to the UV cutoff $k = 2\pi$ (corresponding to a length cutoff at $O(\xi_0)$ in our dimensionless units). In Fig.~\ref{kP}~(a) we plot $kP(k)$ from the numerical simulation for $t> 0$ in multiples of $\sqrt{\tau_Q}$. As we expect, the power gets pushed into longer and longer wavelengths. The scale over which that happens becomes clearer if we plot the moment $I(t)$. In Fig.~\ref{kP}~(b) we compare $I(t)$ measured in the simulation to the analytic behaviour derived from Eq.~(\ref{lgpower2}).
$I(t)$ is seen to stabilise at a value proportional to ${\bar\xi}^{-1}$ by the time defects form.
The Gaussian approximation drives the power to longer wavelengths than the simulation but, as we know from elsewhere~\cite{rivers2001}, the scaling behaviour of the dominant wavelength (corresponding to $\bar\xi^{-1}$) remains the same.

\section{Annular superconductors in three dimensions}

 Even thin annuli are not one-dimensional and we should take the
 annulus width into account. Moreover,
even a
 superconductor that is effectively a 2D film cannot be treated
 entirely as two-dimensional since the three-dimensional nature of the flux
 lines cannot be ignored. In fact, the KZ analysis is potentially
 incomplete, in that the possibility exists of long wavelength modes
 of the electromagnetic field freezing in as the transition is
 implemented. However, we have seen previously (in Ref.~\cite{weir2012}) that this secondary mechanism, the
 Hindmarsh-Rajantie (HR) mechanism~\cite{hindmarsh2000,rajantie2001},
 does not seem to be important for our small systems and we shall not pursue it further.

\subsection{The 2D Gaussian approximation}

For the moment we restrict ourselves to the complex field $\phi_{a}({\bf x})$, extended now to the $x_1-x_2$ plane, assumed Gaussian, as before.  As it stands it is too difficult to get analytic results for arbitrary annuli so, as a first step, we consider a 2D superconducting film of infinite extent and we look for the behaviour of the winding number along a circular closed loop in the film of circumference $C$, that encloses a surface $S$.

If $\rho_3({\bf x})$ is the topological density in the plane~\cite{halperin1981,mazenko}, for given field
configurations $\phi_{a}({\bf x})$ the phase change $\theta_{C}$
along the path can be expressed as the surface integral
\begin{equation}
\theta_{C} = 2\pi\int_{{\bf x}\in S} d^{2}x\,\,\rho_{3}({\bf x})
= -2\pi\int_{{\bf x}\not\in S} d^{2}x\,\,\rho_{3}({\bf x})
\end{equation}
from charge conservation. It then follows
 that, for Gaussian fields, $\langle n^2\rangle$
satisfies
\begin{equation}
\langle n^2\rangle =-\int_{{\bf x}\not\in S}d^{2}x\int_{{\bf
y}\in S}d^{2}y\,\,\langle\rho_3 ({\bf x})\rho_3
({\bf y})\rangle, \label{per}
\end{equation}
where ${\bf x}$ and ${\bf y}$ are in the plane of $S$. Using the results of Refs.~\cite{halperin1981,mazenko} it is not difficult to express the density correlation in terms of the field correlation $f(r)$ ($r = |x|$),
\begin{equation}
\langle\rho_3 ({\bf x})\rho_3
({\bf 0})\rangle = \frac{1}{4\pi^2r}\frac{\partial}{\partial
r}\bigg(\frac{{f^\prime}(r)^2}{1 - f(r)^2}\bigg).
\label{C2}
\end{equation}
It follows from Eq.~(\ref{fshort}) that the density correlation has a range $O(\xi_r)$.
With ${\bf x}$ outside $S$, and ${\bf y}$ inside $S$, all the
contribution to $\langle n^2\rangle$ comes from the vicinity
of the boundary of $S$, rather than the whole area. For large rings this means that, if we
removed all material except for a strip of width $O(\bar\xi )$ from the neighbourhood of the
contour $C$ we would still have the same result. As before, this gives the perimeter rule
\begin{equation}
\langle n^2\rangle\approx a\frac{C}{\xi_r}. \label{per2}
\end{equation}
That is, we have the anticipated random walk in phase along the contour, from which we recover the canonical scaling of Eq.~(\ref{xibar}) on identifying $\xi_r$.

For small annuli the situation is less clear. While  the behaviour of Eq.~(\ref{per}) suggests that the winding number along a small closed path in a 2D superconducting film is proportional to the area of the ring (and hence doubles the KZ exponent), it is not clear that this behaviour would be preserved if the interior of the closed path were removed to make a broad annulus. A test of this is to preserve the perimeter of the annulus but to take different aspect ratios (i.e. different areas). A perimeter law would leave probabilities unchanged, but an area law would lead to differences. The relevance of this is that in an earlier experiment~\cite{monaco2009}, results suggested a doubling of the KZ exponents -- albeit with large errors.

\subsection{Simulations}

To test the approximations outlined in the previous sections we
simulated
the superconducting ring
in a 3D box~\cite{weir2012}.
Although the superconductor is thereby taken to be a thin planar film, the three-dimensional simulation allows for nontrivial correlations of the magnetic field.
Periodic boundary conditions are used. The quench protocol is the same as for the 1D simulations, meaning $\epsilon(t)$ is of the form given in Eq.~(\ref{eq:quench}).

\begin{figure}
\centering
\includegraphics[width=0.75\textwidth]{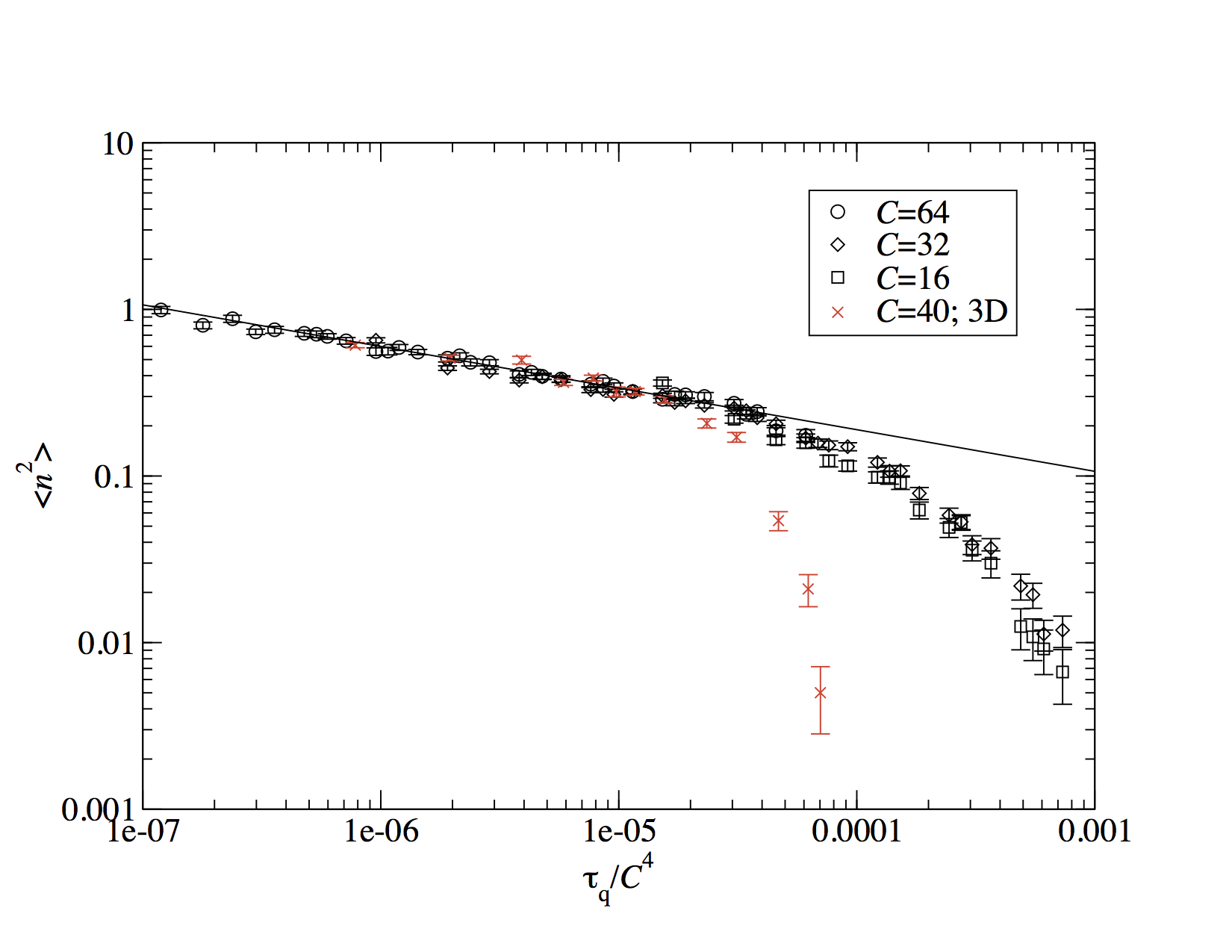}
\caption{\label{fig:volexpsup} Results of simulations for annuli with
  inner circumference $C$ in 1D and 3D, as a function of
  $\tau_Q/C^4$.
In both a change in behaviour from KZ scaling to exponential suppression occurs. We see that there is complete agreement for large annuli (fast quenches) and different exponential damping for small annuli (slow quenches).
}
\end{figure}

We first sought to see if the exponential falloff in trapping
probability predicted above by 1D simulations
was the same for
 the more
realistic theory of 2D rings in three dimensions.   In
Fig.~\ref{fig:volexpsup}, we show the crossover from KZ behaviour to
exponential damping as a function of the combination $\tau_Q/C^4$, for
both the 1D and 3D systems discussed above.
While there is complete agreement
for large annuli the
 exponential damping for small
annuli is dimension dependent. This is
perhaps because of the differences between the
1D expression Eq.~(\ref{napprox}) and the 2D equivalent (\ref{C2}), on
substituting Eq.~(\ref{JTF}) for $f(r)$, even though the annulus is not particularly wide~\cite{kav}. As we would have anticipated, this difference is insensitive to self-coupling strength, as we have checked (but not displayed).

\section{External fields}

All the analysis above assumed no external fields. Experimentally, this is unrealistic.
Moreover, driven symmetry breaking through the application of external bias is an interesting question in its own right. The KZ scenario cannot address this, but we shall find that we can obtain results in the Gaussian approximation that can be analysed in simulations.

Experimentally we compensate for unknown stray fields by applying an external bias field to cancel them. We know this has happened when the likelihood of seeing spontaneous flux is at a minimum. To do this successfully we need to know the likelihood of fluxoid production in the presence of external fields.
It is difficult to extend the previous analysis with periodic boundary conditions to include external fields and we adopt a simpler straightforward variant of the Gaussian approximation.

\subsection{Fluxoid production in an external field: 1D approximation}
\label{sec:prodin1d}

Let us now apply a magnetic flux $\Phi_{f}$ to the loop; this breaks the $\theta\rightarrow -\theta$ symmetry.  The effect of $\Phi_{f}$ is to produce a non-zero average winding number $\langle n\rangle = {\bar n}(\Phi_{f}) = \Phi_{f}/\Phi_0$ after the transition, where $\Phi_0 = hc/2e$.

Because of the limitations of one dimension, the electromagnetic field can only be introduced classically.
The natural extension of our Gaussian probability model~\cite{monaco2008}, $G_N(\Theta)$ in Eq.~(\ref{Ggauss}), for a superconducting loop linked to a magnetic flux $\Phi_{f}$ is that the phase distribution will still be normal with the same variance $\sigma^2(N)$,  but with non-zero average $\bar\Theta(\Phi_{f}) = 2\pi{\bar n}(\Phi_{f})=2\pi\Phi_{f}/\Phi_0=2\pi\phi_{f}$
\begin{equation}
G_{N}(\Theta , \phi_{f}) = \frac{1}{\sqrt{2 \pi
\sigma^2(N)} } \exp-\frac{(\Theta-2\pi\phi_{f})^2}{2
\sigma^2(N)}.
 \label{Ggauss2}
 \end{equation}
With such a distribution, the probability $p_m$ of ending up with a given winding number $m$ is given by
\begin{equation}
p_m(\phi_\mathrm{f}) = \int_{-\pi+2m\pi}^{\pi + 2m\pi} d\theta \; G(\theta;
\phi_\mathrm{f}).
\end{equation}
That is
\begin{equation}
p_{\pm m}(N,\phi_{f})= \frac{1}{2} \bigg[ \textrm{erf} \frac{(\pm 2m- 2\phi_{f} +1)\pi}{\sqrt{2 \sigma^2(N)}} - \textrm{erf} \frac{(\pm 2m -2 \phi_{f} -1)\pi}{\sqrt{2 \sigma^2(N)}}\bigg]
\label{fsfs}
\end{equation}
which, in the presence of a possible stray or residual flux $\phi_r$, we reparametrise as
\begin{equation}
\label{eq:f0}
p_m(\phi_\mathrm{f}) = \frac{1}{2}\left[\mathrm{erf}\,
  \frac{\phi_\mathrm{f} - \phi_\mathrm{r} - m + 0.5}{s_d} - \mathrm{erf}\,\frac{\phi_\mathrm{f} -
    \phi_\mathrm{r} - m -0.5}{s_d}\right],
\end{equation}
where the dependence on quench time and geometry is parameterised by $s_d\equiv\sqrt{2\sigma^2(N)/2\pi}$. Now, $p_{+m}(\phi_{f})\neq p_{-m}(\phi_{f})$. Essentially, what really matters is the difference $m-\phi_f$; for example, $p_{m}(1)=p_{m-1}(0)$. This calculation unfortunately does not allow us to determine the dependence of $s$ on $\tau_\mathrm{Q}$. Nevertheless, we can fit data for different applied magnetic fields at fixed $\tau_\mathrm{Q}$ to Eq.~(\ref{eq:f0}) with only $s_d$ and $\phi_\mathrm{r}$ as free parameters to yield $f_0$.

 We have seen that the Gaussian probability is a good approximation for large rings and, as before, we extrapolate to small rings, using Eq.~(\ref{fsfs}) as it stands.

\section{Experiments in an external field}

\begin{figure}[tb]
\centering
\includegraphics[width=0.3\textwidth]{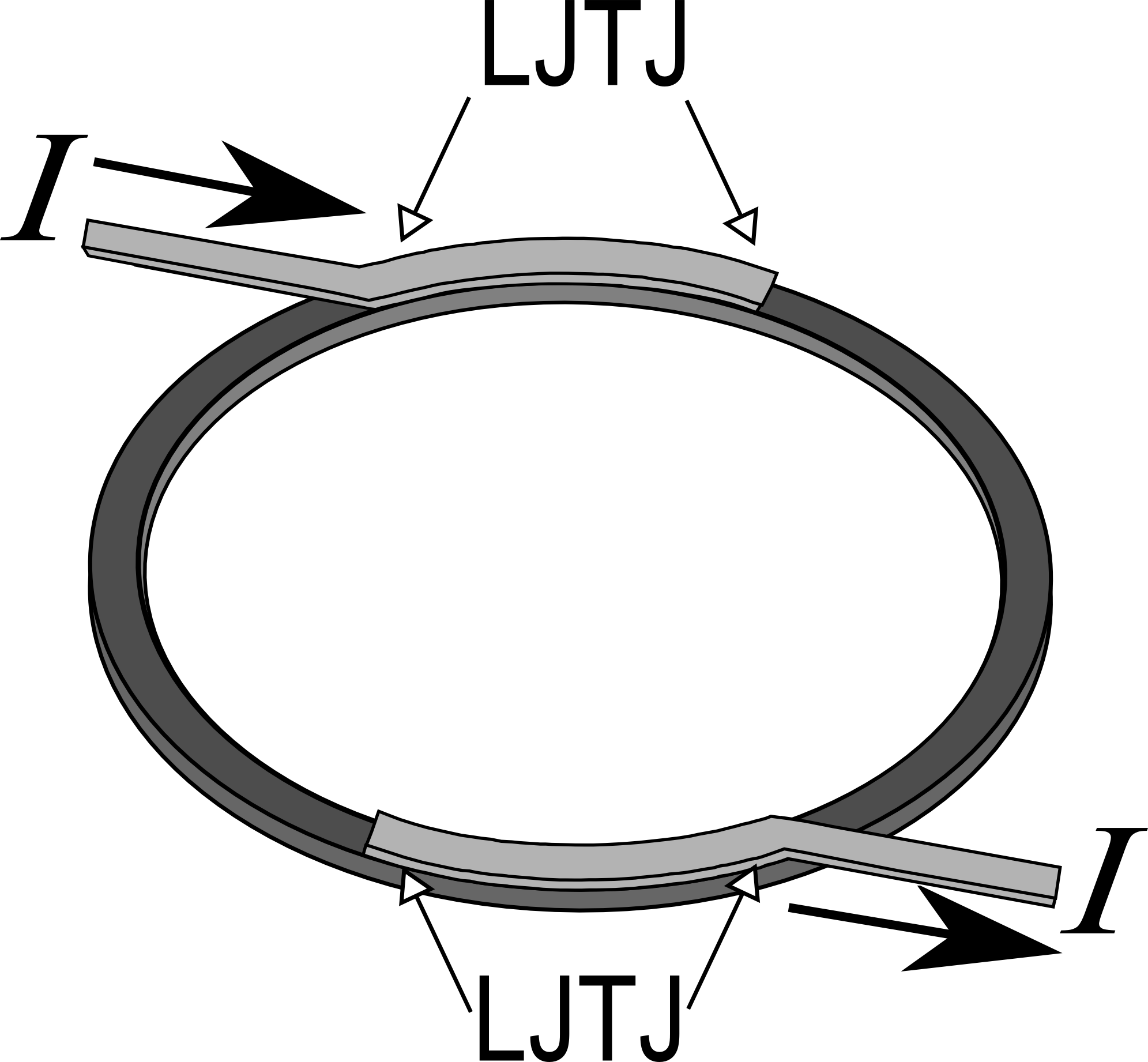}
\caption{The magnetic field associated with the trapped fluxoids can
  be detected by one or more Long Josephson Tunnel Junctions, where the base
  electrode (dark grey) is formed by the ring itself. The top electrodes of the two planar Josephson tunnel junctions are in light gray. \label{ringb}}
\end{figure}

For a thin-film loop, of width $w$ much larger than its thickness but smaller than the ring radius, the intensity of the radial magnetic field $H_\rho$ at the film surface is
\begin{equation}
H_\rho = \frac{I_{cir}}{2w}=\frac{n \Phi_0-\Phi_e}{2wL_{loop}}.
\label{radial1}
\end{equation}
 Eq.~(\ref{radial1}) indicates that a superconducting loop acts as a flux-to-field transformer; if a magnetic sensor is placed above (or below) part of the loop, it will thus detect the local radial field $H_\rho$ and hence the magnetic flux $\Phi_e$ linked to the loop and the winding number, $n$. 
 In this context, the most natural magnetic sensors are planar Josephson tunnel junctions. More specifically, the critical current of Long Josephson Tunnel Junction (LJTJ) -- for which the loop itself constitutes one of the superconducting electrodes -- is able to resolve flux changes well below the flux quantum~\cite{PRB12}.
   Fig.~\ref{ringb} shows a superconducting ring acting as the base electrode for two LJTJs.
Let us stress that the critical current of a LJTJ is sensitive to the surface field, $H_{\rho}$, whereas the superconducting loop is only sensitive to the perpendicular field, $H_{\perp}$.

\begin{figure}[tb]
\centering
\includegraphics[width=0.5\textwidth]{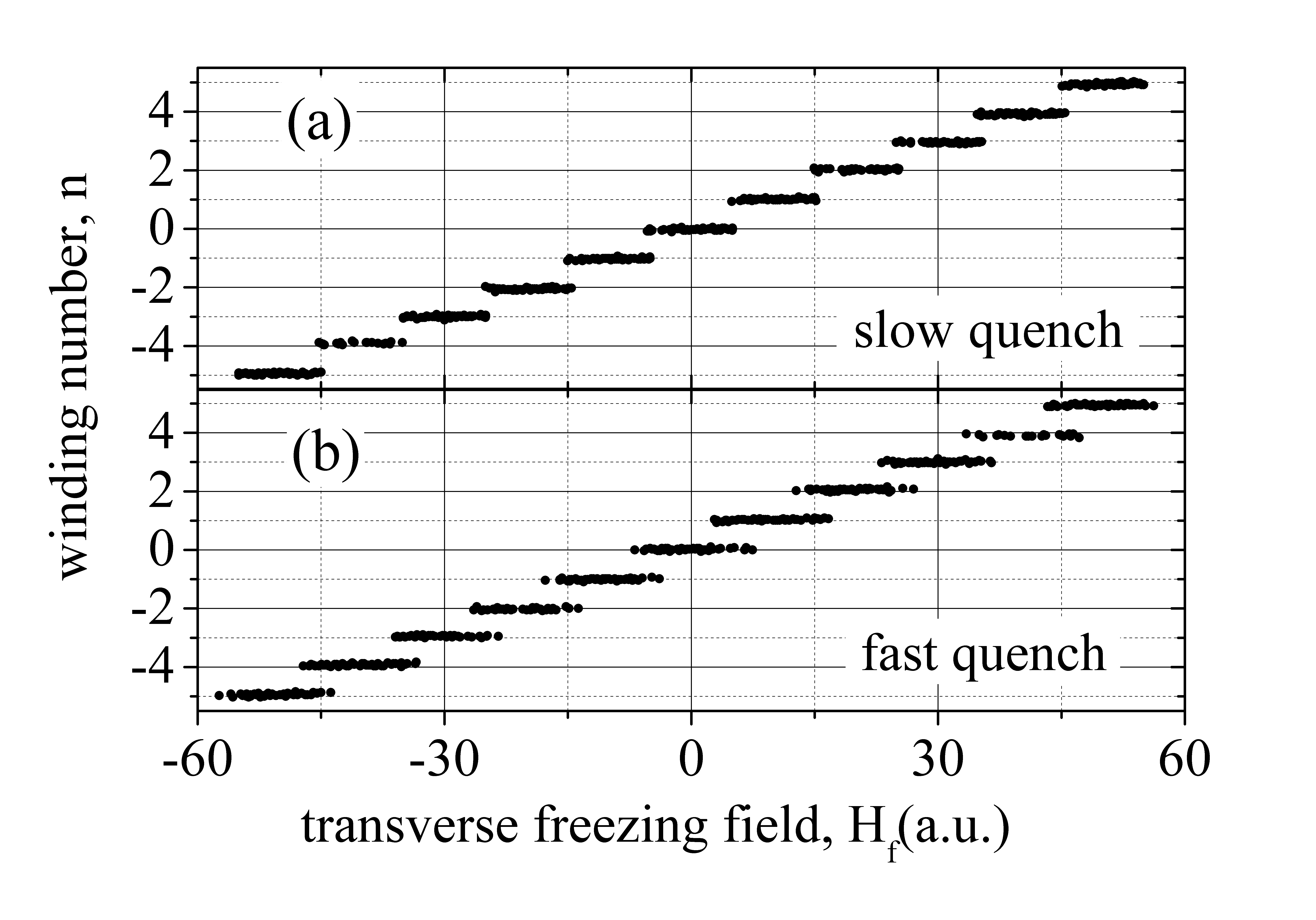}
\caption{Winding number, $n$, vs. the magnetic transverse field,
  $H_f$, with which the loop was field-cooled through its
  superconducting transition temperature, $T_c\approx 9.1\,$K. (a)
  slow quench: the temperature rate change, $dT/dt$, during the
  cooling through the critical temperature was $-0.5\,$K/ms; (b) fast
  quench: $dT/dt \sim -50\,$K/ms. Arbitrary horizontal and vertical
  offsets.
Note that $H_f$ was switched off
  during the readout.}
\label{calib}
\end{figure}

Figs.~\ref{calib}~(a) and (b) show the quantum levels observed when the system is cooled through a controlled Normal-Superconducting (NS) transition in the presence of a transverse magnetic field, $H_f$, which is incremented by steps corresponding to $\sim 0.02\,\Phi_0$ jumps in the freezing flux, $\Phi_f=\mu_0 H_f A_{loop}$, where $A_{loop}$ is the loop area. Once $H_f$ is removed, each trapped flux quantum results in a small change in $H_{\rho}$ which, in turn, produces a detectable change in the junction zero-field critical current, $I_c$. The two panels of Fig.~\ref{calib} refer to the same sample quenched in the same field range, but with different cooling rates, $dT/dt$: in the first panel it is about $-500\,$K/s, while it is hundred times larger for the second panel where a marked overlap of the quantum states is visible.
In deriving Figs.~\ref{calib}~(a) and (b), we assumed that the (unknown) residual magnetic field of our setup was small enough to be trapped in the loop as less than one fluxoid. Nevertheless, the presence of any larger (static) magnetic field would simply result in a horizontal shift of the field axis or, correspondingly, a vertical shift of the winding number, $n$. In other words, we are able to measure the changes in the loop quantum state, but not its absolute value. Luckily, as discussed in Section~\ref{sec:prodin1d}, the transition from the $n$-th to the $n+1$-th state is virtually independent of $n$.

\subsection{Equipment used}

Our setup consisted of a cryoprobe inserted vertically in a
commercial \textsl{LHe} dewar. The cryoprobe was magnetically shielded
by means of two concentric \textsl{Pb} cans and a vacuum tight
cryoperm can surrounding them and immersed in the \textsl{LHe} bath
($T\simeq 4.2\,$K). In addition, the measurements were carried out in
an RF-shielded environment.
We used high quality
all-Niobium LJTJs fabricated on $4.2 \times 3\,$mm$^2$ silicon
substrates using the trilayer technique;
the Josephson
junction is realized as a window opened in a
\textsl{Si$\mathrm{\textsl{O}}_2$} insulator layer.
The samples'
parameters can be found in
Ref.~\cite{PRB12}.

The chip was mounted on
a massive
\textsl{Cu} block. Inside the outer can
\textsl{He} exchange gas
with a pressure of about $20\,$mbar provided the thermal link between
the \textsl{Cu} block and the \textsl{LHe} bath. The chip was heated
above the loop critical temperature, $T_c\approx 9.1\,$K, by a laser
pulse transmitted along an optical fiber
to the back side of
the chip. The single crystalline \textsl{Si} chip absorbed a large
fraction of the incident green light, and its very high thermal
conductance minimized thermal gradients.
After the
laser pulse,
excess heat diffuses away from the chip through the
thermal contact with the \textsl{Cu} block and the \textsl{He} gas
inside the can; the chip temperature then relaxes down to the bath
temperature.

During the transition from the normal to the superconducting states, a
calibrated magnetic field $H_{\perp}$ was applied
perpendicular to the loop plane by means of a superconducting
cylindrical coil aligned with the loop axis, while the LJTJ was
electrically isolated; in fact, both the junction voltage and current
leads were open during the whole thermal cycle. At the end of each
cycle, the transverse field was removed and -- as previously explained
-- the possible fluxoids are counted by a measurement of the LJTJ's
zero-field critical currents.
This method works well as
long as the probability $f_2$ of trapping two fluxoids is small. Hundreds
of thermal cycles were  carried out for each value of the trapping
field, $H_{\perp}$; the field value was increased in steps
of
about $0.1-0.2\Phi_0$
until a total flux
variation, $\Delta \Phi_f$, well above one flux quantum was
achieved. In order to run batches of several thousands of equal
thermal cycles with given constant parameters, the automation of
thermal cycles was implemented by means of a switching unit controlled
by a GPIB interface; this also allowed for
more robust statistics
to be achieved. At the end of each thermal quench, the zero-field
junction current-voltage characteristic is automatically digitally
acquired and stored.

\begin{figure}[tb]
\centering
\includegraphics[width=0.5\textwidth]{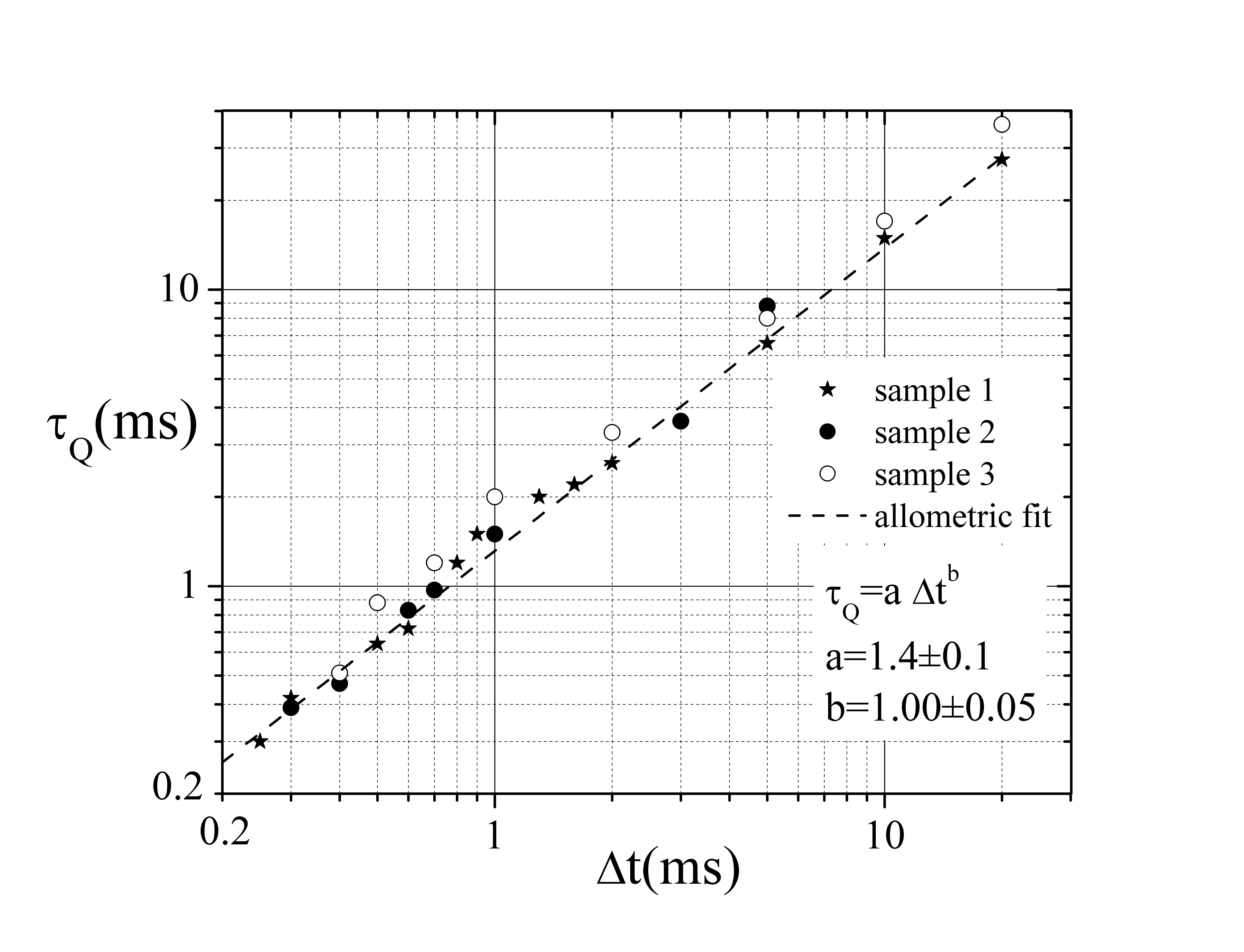}
\caption{Quenching time, $\tau_Q$, vs the duration, $\Delta t$, of the
  laser pulse for three samples. The dashed line shows the best
  allometric fit. \label{tauq}}
\end{figure}

The temperature dependence of the LJTJ's gap voltage was used as an on-chip fast thermometer, to monitor the chip temperature during the thermal cycle.
The quench time, $\tau_Q$, was inferred from a well known fitting
procedure~\cite{monaco2006b}. Fig.~\ref{tauq} shows that the quench time $\tau_Q$ is proportional to the laser pulse duration $\Delta t$, with the proportionality constant almost independent of the sample (and of its mounting procedure). Indeed, this method for reading the loop winding number
using
a LJTJ was
adopted in a previous work~\cite{monaco2009}, but it became reliable and efficient only when the $\delta$-biased~\cite{PRB10} junction was replaced by an in-line one~\cite{PRB12,ferrel,os}, as
shown in Fig.~\ref{ringb}.

\subsection{Fluxoid readout}

\begin{figure}[tb]
\centering
\includegraphics[width=0.4\textwidth]{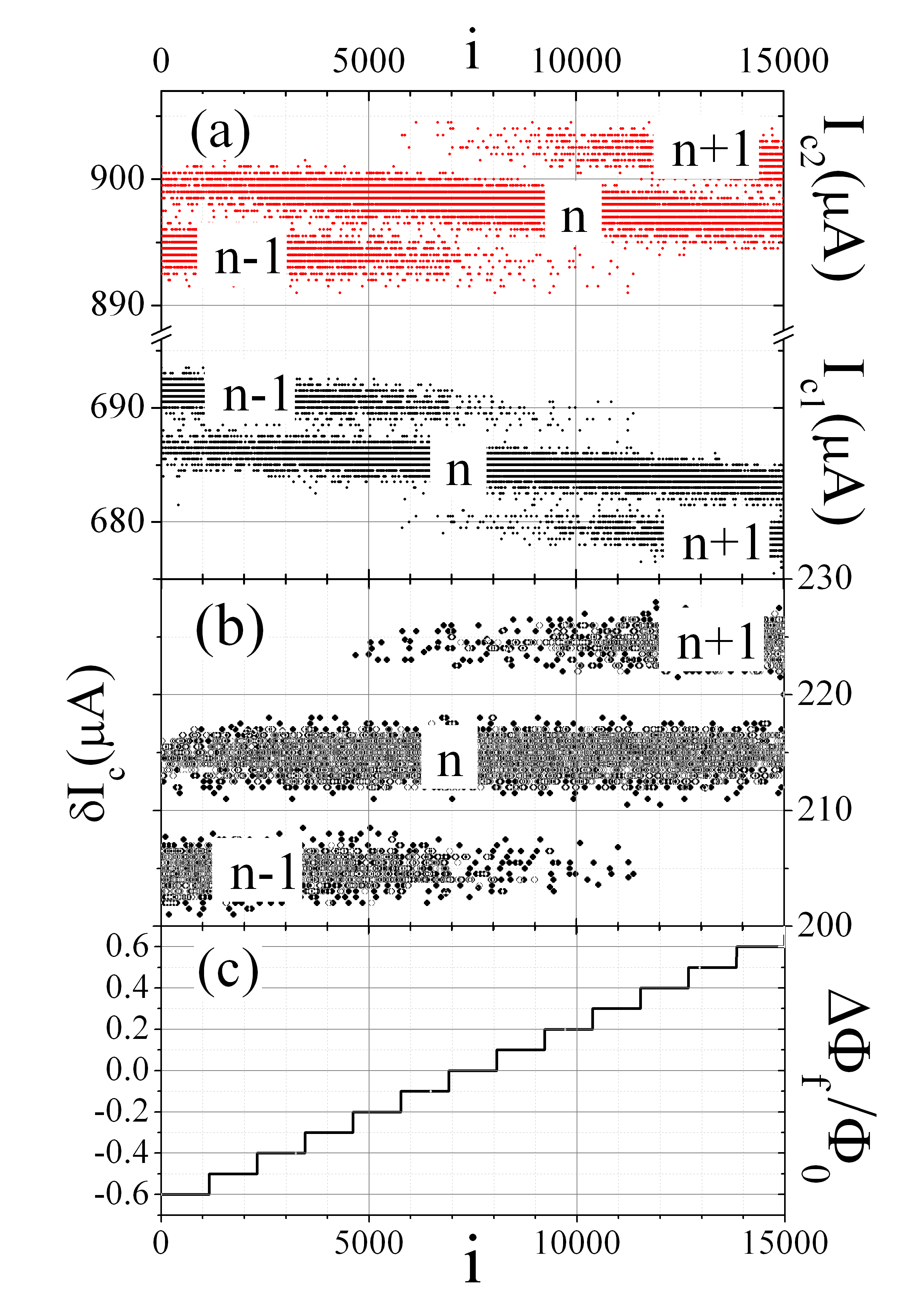}
\caption{ (a) Zero-field critical currents $I_{c1}$ (dark points) and $I_{c2}$ (light points) as a function of the cycle number $i$. (b) The same for $\delta I_c = I_{c2}-I_{c1}$. (c) Relative variation of the freezing flux, $\Phi_f$. All plots have an arbitrary vertical offset.
\label{ics}}
\end{figure}

The discrete variation $\Delta I_c$ of the zero-field critical current associated with each flux quantum trapped in the loop is~\cite{PRB12}
\begin{equation}
\Delta I_c=g \frac{\Phi_0}{L_{loop}},
\label{Delta}
\end{equation}
where $g$ is a dimensionless parameter of the order of unity that depends on some geometrical details~\cite{PRB12}. With a loop inductance $L_{loop}\approx 100\, $pH, Eq.~(\ref{Delta}) provides $I_c$ jumps of the order of several microamperes in a flux range of hundreds of flux quanta~\cite{SUST12}. The discreteness of the critical current values is wider when the loop
has
a top junction electrode
that is narrower and thicker than the bottom one, resulting
in a larger gain factor $g$~\cite{SUST13}. The validity of Eq.~(\ref{Delta})
is
supported by measurements on a number of devices based on narrow loops with annular and rectangular geometries.

\subsection{Common mode}

In the common mode configuration,
shown in
Fig.~\ref{ringb}, the two LJTJs share the same doubly connected base
electrode and are series biased. Any circulating current, $I_{cir}$,
modulates the zero-field critical currents, $I_{c1}$ and
$I_{c2}$, of the two LJTJs with opposite sign. As a result their difference, $\delta I_c =
I_{c2}-I_{c1}$, changes twice as fast as the winding number,
$n$.
The rms current noise on $\delta I_c$ is only $\sqrt{2}$
times larger than that on a single critical current, meaning that in
the common mode configuration the signal-to-noise ratio is enhanced by
a factor $\sqrt{2}$.
Furthermore,
by reading the two critical currents simultaneously and
looking at
their correlation,
it is possible to
exclude unwanted events from the statistical analysis.
These are related to the trapping of Abrikosov vortices found at
pinning centres of the superconducting electrodes.
A further reduction of the signal-to-noise ratio by a factor of $\sqrt{2}$
is achieved by taking the combination, $I_{c2}^+ - I_{c2}^- -I_{c1}^+ + I_{c1}^-$, where $I_{c1,2}^+$ and $I_{c1,2}^-$ are the absolute values of, respectively, the positive and negative critical currents of the two LJTJs.

\subsection{Measuring the trapping frequency}

\noindent Fig.~\ref{ics}(a) shows the zero-field critical currents,
$I_{c1}$ and $I_{c2}$, of a two-junction annular sample during a
statistical batch in which each value of the cooling field,
$H_{\perp}$, was maintained for about a thousand thermal cycles.
Since each cycle lasts slightly less
that 4 seconds,
the set of 15000 cycles
required
about twelve hours. The reduction of the critical currents observed
during the measurements corresponds to a small decrease (drift) of the
bath temperature.
Next, Fig.~\ref{ics}(b) plots the
difference $\delta I_c=I_{c2}-I_{c1}$; we observe that the current
jumps have doubled, the signal-to-noise
is lower
and,
in addition,
the effect of
temperature drift has been
counterbalanced. As Fig.~\ref{ics}(c) indicates, the cycle number $i$
shows the step-like dependence of the freezing flux, $\Phi_{f}$, as
time goes by. Let us call $\Phi_r$ the residual magnetic flux;
then, to reproduce the zero-flux condition during the quench, a
freezing flux $\Phi_f = -\Phi_r$ needs to be applied. Since $\Phi_r$
is, in general, unknown, it is necessary to span $\Phi_f$ over a range at least as large as one flux quantum.

\begin{figure}[tb]
\centering
\subfigure[]{\begin{minipage}[c]{0.45\textwidth}
\centering
\includegraphics[width=\textwidth]{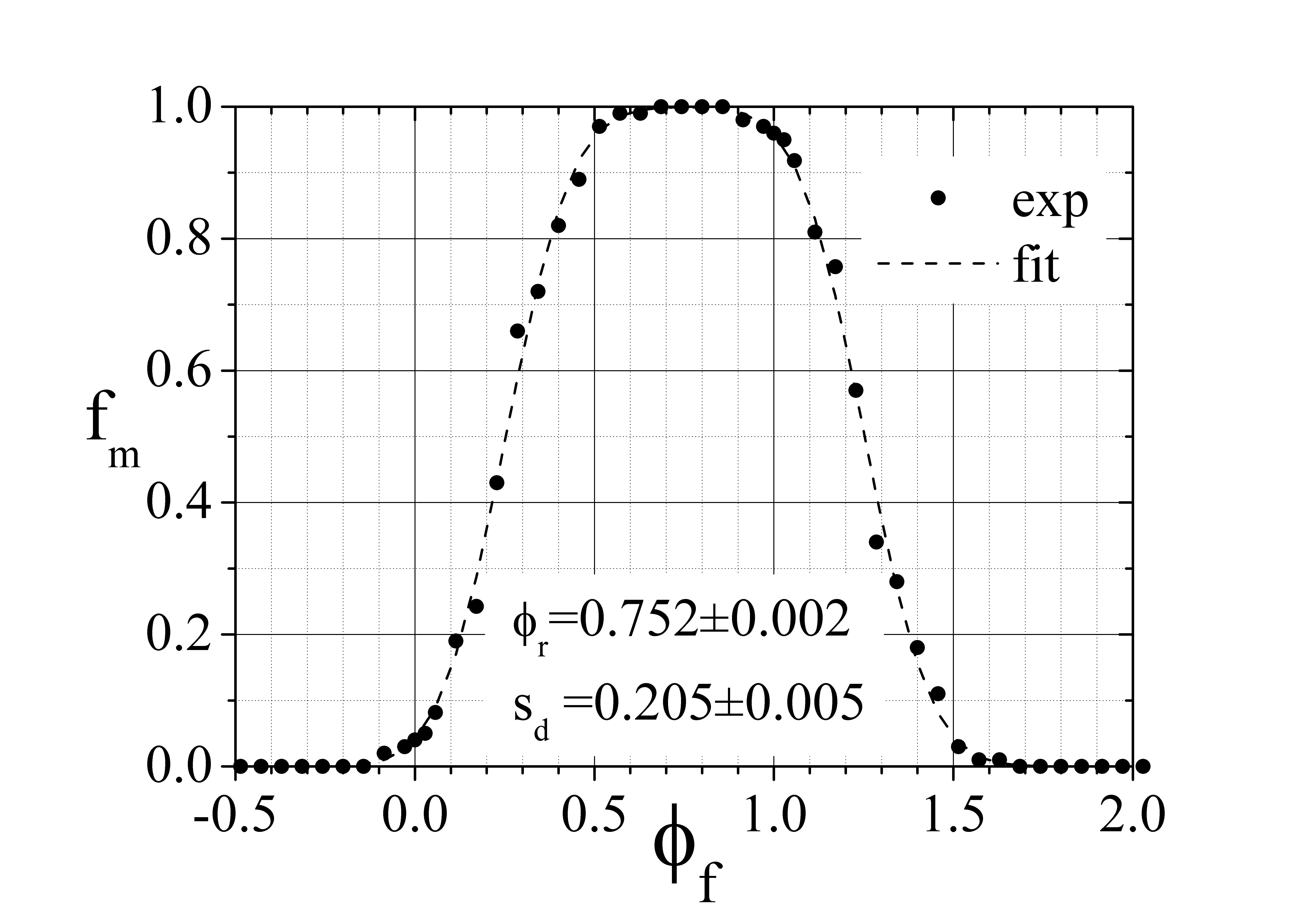}
\end{minipage}}
\subfigure[]{\begin{minipage}[c]{0.45\textwidth}
\centering
\includegraphics[width=\textwidth]{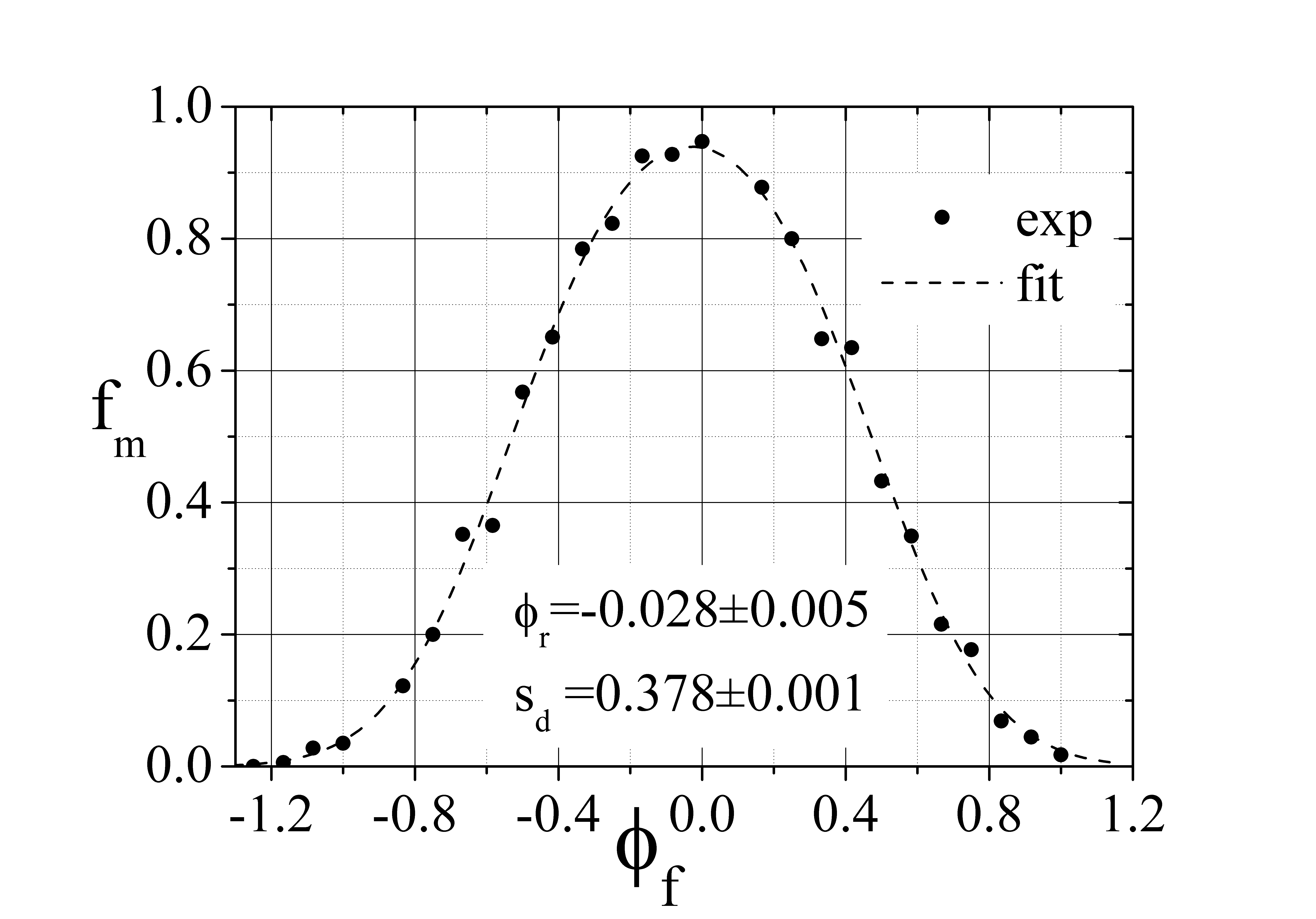}
\end{minipage}}
\caption{Experimental probability ${\bar p}_1$ of observing one defect as a function of normalised external magnetic flux $\phi_f$. The curves are a fit of the form given in Eq.~(\ref{eq:f0}).}
\label{distributions}
\end{figure}

\begin{figure}[tb]
\centering
\subfigure[]{\begin{minipage}[c]{0.45\textwidth}
\centering
\includegraphics[width=0.9\textwidth]{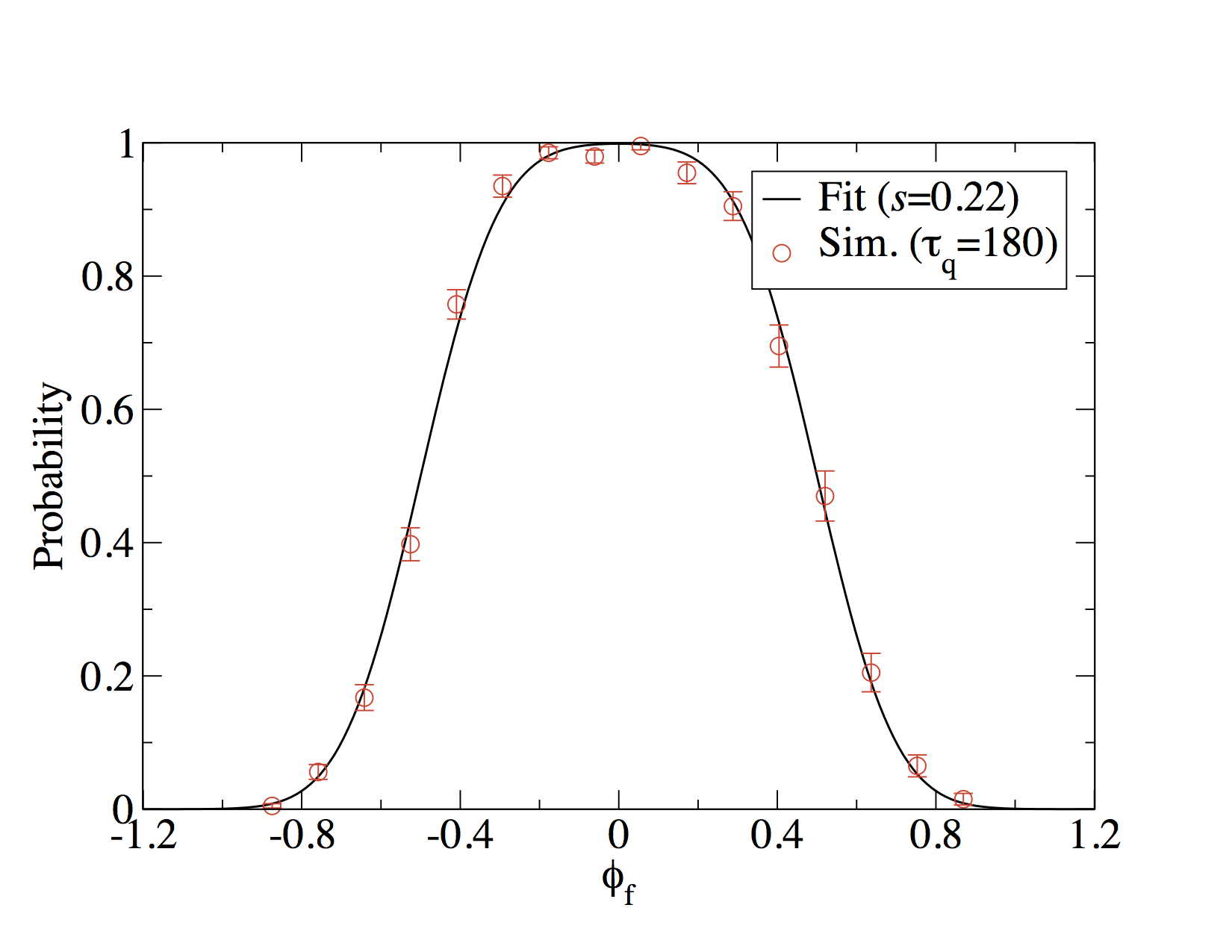}
\end{minipage}}
\subfigure[]{\begin{minipage}[c]{0.45\textwidth}
\centering
\includegraphics[width=0.9\textwidth]{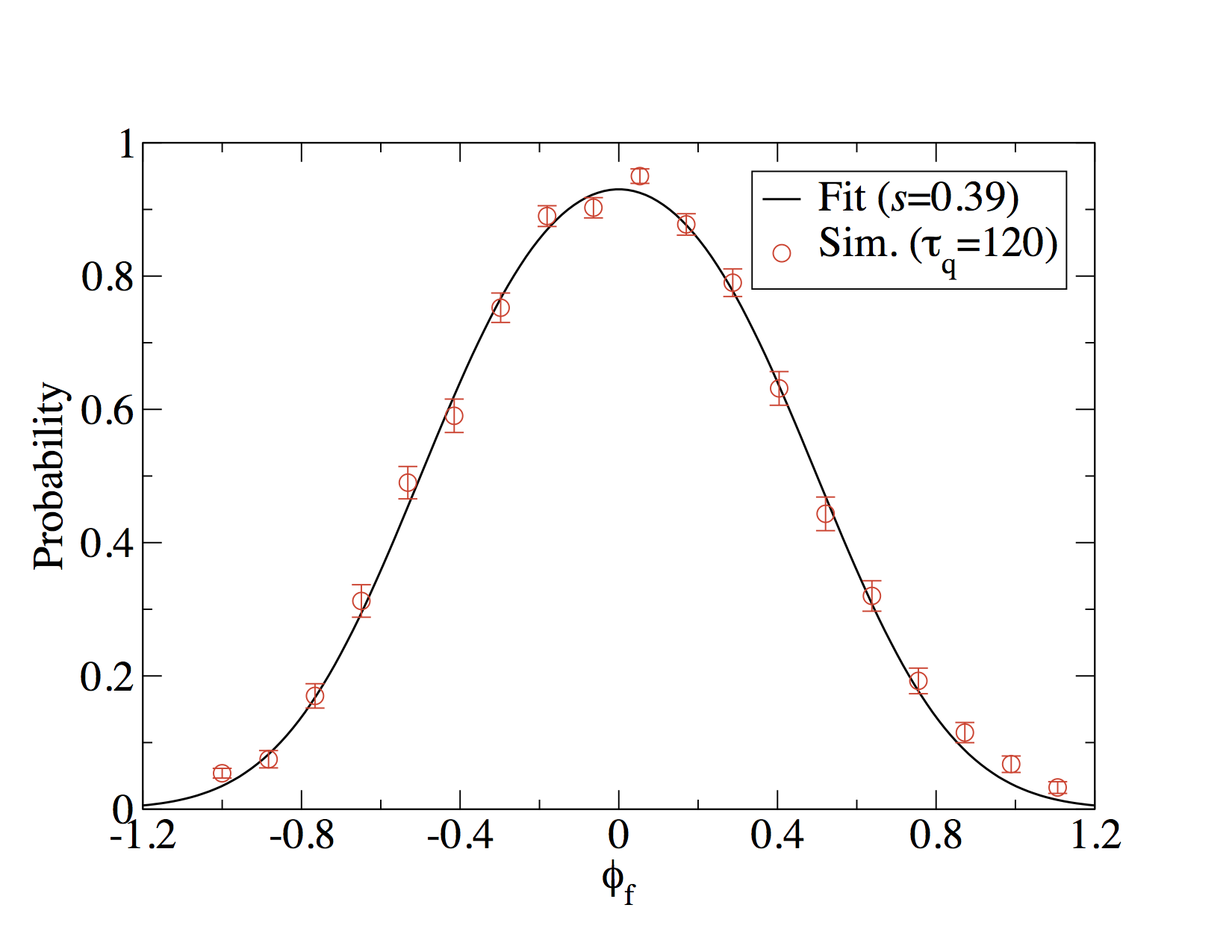}
\end{minipage}}
\caption{As Fig.~\ref{distributions} but for 3D simulations with a nonzero external field. The peak probabilities are comparable.
The experimental measurements, the simulations and the Gaussian approximation all agree.}
\label{simdists}
\end{figure}

From the statistical analysis of data thus obtained
it is possible to determine the frequency distributions, $f_m$.
Two examples of the frequency dependency on the external field are given in Figs.~\ref{distributions}~(a) and (b);
we additionally plot 3D simulations for comparable zero-field fluxoid production frequencies in Figs.~\ref{simdists}. The data are very nicely fit by the Gaussian form of Eq.~(\ref{eq:f0}).

Unfortunately, we do not have enough
experimental data to show how $\bar p_1$ varies with $\tau_Q$, except that it is in the exponentially suppressed regime.

\section{Conclusions}

In this paper we have explored the ways in which spontaneous fluxoid production in annular superconductors can
cast light on the Kibble-Zurek (KZ) mechanism in systems that are far from equilibrium.

The simple picture is complicated by two effects: small size and symmetry-breaking external fields. The KZ scenario is unable to address either of these. However, we know that the basic scaling result Eq.~(\ref{xibar}) also provides a good test of the Gaussian behaviour of the order parameter field at the time of fluxoid production, in a way that correctly encodes the KZ scales. We have extended the Gaussian approximation to those areas which the KZ scenario cannot reach.

As it stands, the Gaussian approximation is only simply analytic for an
idealised 1D annulus, in which we can include an external magnetic
field classically. In fact, we have adopted two variants of
Gaussianity which differ very slightly: one is more convenient for
addressing periodicity; the other for including external fields.
We have performed more simulations based on TDGL theory, both for
the
idealised 1D annulus and
the
more realistic annulus with finite width embedded in 3D space.
Where comparable,
we find no important difference between the 1D and 3D simulations.

For large rings we reproduce the canonical KZ scaling.
However, the likelihood of seeing fluxoids for small ring size or slow quenches is exponentially damped in a dimensionally dependent way.

In the presence of the explicit symmetry breaking of external fields we find strong agreement between the 3D simulations and the 1D results of the Gaussian approximation.
Indeed,
experimental results for \textsl{Nb} annuli agree totally with both.
The implications for the KZ scenario are strong. Rather than think in
terms of causal horizons as the constraints on growth of correlations,
we see field ordering in terms of the fastest possible growth of
unstable long wavelength modes, while encoding KZ scales.

We finish with an important issue that
requires
further study.
This is the empirical
observation of scaling behaviour in this, and other systems, for which the observed exponents
initially take {\it double} the canonical KZ values, as seemed to be
the case in our earlier experiment on superconducting annuli~\cite{monaco2009}.  Further, a
recent experiment on defect production in linear ionic crystals again
shows twice the expected exponent~\cite{delcampo2012}. We believe
there are different reasons for these. In Ref.~\cite{monaco2009} we
provided a possible Gaussian explanation for superconductors in terms
of small size effects, which
was applied
by the authors
of Ref.~\cite{delcampo2012}. Preliminary analyses here suggest something
more complicated.
The conditions
imposed in
Ref.~\cite{monaco2009} for doubling in 1D systems are not satisfied for
superconducting loops. Unless there is an effect due to annulus width
(which Gaussianity suggests for wide annuli but which we have been
unable to reproduce in simulations) our `doubling' of Ref.~\cite{monaco2009} -- with its large errors in comparison to those of Refs.~\cite{monaco2006a,monaco2006b} --
is most likely the shoulder in Fig.~\ref{fig:volexpsup} as the exponential damping takes hold. Indeed, recent experiments involving Josephson vortices in Bose-Einstein condensates show similar exponential behaviour~\cite{collapse}.

The situation is different for short linear systems with open boundary conditions for double-well $Z_2$ -- rather than $U(1)$ -- symmetry breaking, where we do seem to see doubling of the exponent in simulations (not presented here). This is more
relevant
to linear ionic crystals, giving a possible way to understand empirical exponent doubling seen there~\cite{delcampo2012}.

The doubling of KZ scaling exponents for fluxon production in annular Josephson tunnel Junctions (presented in Refs.~\cite{monaco2006a,monaco2006b}) is the most complicated of all.
On the one hand, it may be due to the fabrication of the junction, a proximity effect arising from the
deposition of non-superconducting metal on the oxide interface. There is possibly another cause,
the presence of pinning centres for vortices.
Considering a short Josephson tunnel junction as an idealisation of a narrow annular junction with a strong pinning centre,
an analogous
doubling effect
to that
seen in $Z_2$ symmetry breaking may take place~\cite{ana}. This is under study.

\ack
The authors wish to thank Arttu Rajantie for helpful discussions. RR
thanks the Helsinki Institute of Physics for hospitality, where some
of this work was performed. The simulations were carried out using the
resources of the Imperial College High Performance Computing Service.

\end{document}